\def\e{{\rm e}}
\def\del{\partial}

\def\abs#1{{\left|{#1}\right|}}
\def\vev#1{\langle #1 \rangle}
\def\del{\partial}

\def\abs#1{{\left|{#1}\right|}}
\def\vev#1{\langle #1 \rangle}

\def\del{\partial}
\def\dslash{\del\kern-0.55em\raise 0.14ex\hbox{/}}

\def\rough#1{\raise.3ex\hbox{$#1$\kern-.75em\lower1ex\hbox{$\sim$}}}

\newcommand{\PRD}[3]{{\it Phys. Rev.} {\bf D{#1}} (19{#3}) {#2}}
\newcommand{\PRDO}[3]{{\it Phys. Rev.} {\bf {#1}} (19{#3}) {#2}}
\newcommand{\PRDM}[3]{{\it Phys. Rev.} {\bf D{#1}} (20{#3}) {#2}}

\newcommand{\PRLM}[3]{{\it Phys. Rev. Lett.} {\bf {#1}} {#2} (20{#3})}
\newcommand{\NPB}[3]{{\it Nucl. Phys.} {\bf B{#1}} {#2} (19{#3})}
\newcommand{\NPBM}[3]{{\it Nucl. Phys.} {\bf B{#1}} (20{#2}) {#3}}
\newcommand{\PLB}[3]{{\it Phys. Lett.} {\bf {#1}B} (19{#3}) {#2}}
\newcommand{\PLBM}[3]{{\it Phys. Lett.} {\bf B{#1}} (20{#3}) {#2}}

\newcommand{\PTP}[3]{{\it Prog. Theor. Phys.} {\bf {#1}} (19{#3}) {#2}}

\newcommand{\ANN}[3]{{\it Ann. Phys. (N.Y.)} {\bf {#1}}, {#2} (19{#3})}

\newcommand{\MPL}[3]{{\it Mod. Phys. Lett.} {\bf A{#1}} (19{#3}) {#2}}
\newcommand{\MPLM}[3]{{\it Mod. Phys. Lett.} {\bf A{#1}} (20{#3}) {#2}}

\newcommand{\hmu}{\hat\mu}

\documentclass[12pt]{article}
\usepackage{graphicx}
\begin{document}
\baselineskip=18pt
\begin{titlepage}
\begin{flushright}
TU-788
\end{flushright}
\vspace{1cm}
\begin{center}{\Large\bf 
Large Gauge Hierarchy in Gauge-Higgs Unification}
\end{center}
\vspace{1cm}
\begin{center}
%
%
Kazunori Takenaga$^{(a)}$
\footnote{E-mail: takenaga@tuhep.phys.tohoku.ac.jp
\\Collaboration with M. Sakamoto (Kobe), dragon@kobe-u.ac.jp\\
Talk given at the SCGT06, International 
Workshop on "Origin of Mass and Strong Coupling Gauge 
Theories" 21-24 November 2006, Nagoya, Japan
}
\end{center}
\vspace{0.2cm}
\begin{center}
%
%
${}^{(a)}$ {\it Department of Physics, Tohoku University, 
Sendai 980-8578, Japan}
\end{center}
\vspace{1cm}
\begin{abstract}
We study a five dimensional nonsupersymmetric 
$SU(3)$ gauge theory compactified on $M^4\times S^1/Z_2$. 
The gauge hierarchy is discussed in the scenario of the
gauge-Higgs unification. We present two models in which the 
large gauge hierarchy is realized, that is, the weak 
scale is naturally is obtained from an unique large 
scale such as a GUT and the Planck scale. We also 
study the Higgs mass in each model. 
\end{abstract}
\begin{center}
{\it keywords}:~gauge symmetry breaking, extra 
dimensions, boundary conditions
\end{center}
\end{titlepage}
\newpage
\section{Introduction}
Higher dimensional gauge theory has been paid much 
attention as a new approach to overcome the hierarchy 
problem in the standard model without introducing 
supersymmetry. In particular, the gauge-Higgs 
unification is a very attractive 
idea \cite{manton,hosotani,gaugehiggs1}. The 
higher dimensional gauge symmetry plays a role to 
suppress the ultraviolet effect on the Higgs mass. 
The Higgs self interaction is understood as part 
of the original five dimensional gauge coupling, so 
that the mass and the interaction can be predicted 
in the gauge-Higgs unification. The gauge-Higgs 
unification has been studied 
extensively \cite{gaugehiggs2}.

In the gauge-Higgs unification, the Higgs field 
corresponds to the Wilson line phase, which is nonlocal 
quantity. The Higgs potential is generated at the 
one-loop level after the compactification. Because of
the nonlocality, the Higgs potential never suffers 
from the ultraviolet effect \cite{masiero}, which is 
the genuine local effect, and the Higgs mass
calculated from the potential is finite as well. In 
other words, the Higgs potential and the mass are 
calculable in the gauge-Higgs unification. This is a 
remarkable feature rarely happens in the usual 
quantum field theory. It is understood that the 
feature entirely comes from shift symmetry manifest 
through the Wilson line phase, which is a remnant 
of the higher dimensional gauge symmetry appeared in four
dimensions. The Higgs mass does not depend on the cutoff 
at all, so that two tremendously separated energy 
scales can be stable in the gauge-Higgs unification.

We study a five dimensional nonsupersymmetric $SU(3)$ 
gauge theory, where one of spatial coordinates 
compactified on an orbifold $S^1/Z_2$. We find two 
models (model I, II) which realize the large gauge 
hierarchy \cite{sakatake}.

\section{Gauge-Higgs unification}
As the simplest example 
of the gauge-Higgs unification, we study a 
nonsupersymmetric $SU(3)$ gauge theory 
on $M^4\times S^1/Z_2$, where $M^4$ is the 
four dimensional Minkowski space-time and $S^1/Z_2$ is 
an orbifold which has two fixed points, $y=0, \pi R$. 

We impose the twisted boundary condition of the field 
for the $S^1$ direction and at the fixed points by using 
the gauge degrees of freedom,
\begin{eqnarray}
A_{\hmu}(x, y + 2\pi R) &=&
U A_{\hmu}(x, y ) \, U^\dagger ,
\label{shiki1}
\\
\left(\begin{array}{c}
A_\mu \\
A_y
\end{array}\right) (x, y_i - y) &=& P_i
\left(\begin{array}{c}
A_\mu \\
- A_y 
\end{array}
\right)(x, y_i + y),
P_i^\dagger,~~ (i = 0, 1)
\label{shiki2}
\end{eqnarray}
where $U^\dagger = U^{-1}, P_i^\dagger = P_i= P_i^{-1}$ and
$y_0=0, y_1=\pi R$ and $\hmu$ stands 
for $\hmu=(\mu, y)$. The minus sign for $A_y$ is 
needed to preserve the invariance of the 
Lagrangian under these transformations. A 
transformation $\pi R+y \stackrel{P_1}{\rightarrow}
\pi R -y$ must be the same as $\pi R +y \stackrel{P_0}
{\rightarrow}-(\pi R + y)\stackrel{U}{\rightarrow} \pi R -y$,
so that we obtain $U = P_1 P_0$. Here we 
choose $P_0=P_1={\rm diag.}(-1, -1, 1)$.

The gauge symmetry at low energies consists 
of the zero modes for $A_{\mu}^{(0) a=1,2,3, 8}$. We 
see that the orbifolding boundary condition $P_i$ breaks 
the original gauge symmetry $SU(3)$ down 
to $SU(2)\times U(1)$ at the fixed 
points \cite{orbi}. On the other hand, we observe that 
the zero mode for $A_y^{(0)}$ transforms 
as an $SU(2)$ doublet, so that we identify the Higgs field as
\begin{equation}
\Phi\equiv \sqrt{2\pi R}~\frac{1}{\sqrt{2}}~
\left(
\begin{array}{c}
A_y^{(0)4} -i A_y^{(0)5} \cr 
A_y^{(0)6} -i A_y^{(0)7} \cr
\end{array}\right).
\label{shiki6}
\end{equation}
The VEV of the Higgs field is parametrized, by 
using the $SU(2)\times U(1)$ gauge degrees 
of freedom, as 
\begin{equation}
\vev{A_y^{(0)}}\equiv \frac{a}{g_4 R}
\frac{\lambda^6}{2}=A_y^{(0)6}\frac{\lambda^6}{2},
\label{shiki8}
\end{equation}
where $a$ is a dimensionless parameter. In order to 
determine $a$, one usually valuates the effective
potential for $a$ \cite{hosotani}. 
The gauge symmetry breaking depends on the 
values of $a_0$. It has been known that the 
matter content is crucial 
for the correct gauge symmetry 
breaking $SU(2)\times U(1)\rightarrow U(1)_{em}$.

\section{Large gauge hierarchy in the gauge-Higgs unification} 
If the Higgs acquires the VEV, the $W$-boson becomes 
massive whose mass is given by 
$M_W=a_0/2R$. This relation defines an important ratio,
\begin{equation}
\frac{M_W}{M_c}=\pi a_0,
\label{shiki16} 
\end{equation}
where $M_c\equiv (2\pi R)^{-1}$. Once the values 
of $a_0$ is determined as the minimum of the 
effective potential, the compactification scale
$M_c$ is fixed through Eq.(\ref{shiki16}). In 
the usual scenario of the gauge-Higgs unification, the 
VEV is of order of $O(10^{-2})$ for appropriate 
choice of the flavor set\cite{hty2} and this 
yields $M_c\sim$ a few TeV. In order to realize 
the large gauge hierarchy such 
as $M_c\sim M_{GUT}, M_{Planck}$, one needs the 
very small values of $a_0$.

For the very small values of $a$, the effective potential
can be expanded as 
\begin{equation}
{\bar V}_{eff}(a)= -\frac{\zeta(3)}{2}C^{(2)}(\pi a)^2
+\frac{(\pi a)^4}{24}
\left[C^{(3)}\left(-{\rm ln}(\pi a) +\frac{25}{12}
\right)+C^{(4)}({\rm ln}2)
\right]+\cdots,
\label{shiki19}
\end{equation}
where $V_{eff}(a)\equiv C{\bar V}_{eff}(a)$  with 
$C\equiv {\Gamma({\frac{5}{2}})/{\pi^{\frac{5}{2}}
(2\pi R)^5}}$, and the coefficient $C^{(i)} (i=2,3,4)$ is 
defined by 
\begin{eqnarray}
C^{(2)}&\equiv & 24N_{adj}^{(+)}+4N_{fd}^{(+)}
+\frac{9d}{2}N_{adj}^{(-)s}
+\frac{3}{2}N_{fd}^{(-)s}\nonumber\\
&&-\left(18+6dN_{adj}^{(+)s}+2N_{fd}^{(+)s}
+18N_{adj}^{(-)}+3N_{fd}^{(-)}
\right),\label{shiki20}\\
C^{(3)}&\equiv & 72N_{adj}^{(+)}+4N_{fd}^{(+)}
-\left(54+18dN_{adj}^{(+)s}+2N_{fd}^{(+)s}\right),
\label{shiki21}\\
C^{(4)}&\equiv & 48+16dN_{adj}^{(+)s}
+18dN_{adj}^{(-)s}+2N_{fd}^{(-)s}\nonumber\\
&&-\left(64N_{adj}^{(+)}+4N_{fd}^{(-)}
+72N_{adj}^{(-)}\right).
\label{shiki22}
\end{eqnarray}
We emphasize that each coefficient in the 
effective potential is given  by the discrete 
values, that is, the flavor number of the massless 
bulk matter. This is the very curious feature of the 
Higgs potential, which is hardly seen in the usual 
quantum field theory, and is a key point to
discuss the large gauge hierarchy in 
the gauge-Higgs unification.
\subsection{Model I}
We impose a condition $C^{(2)}=0$ in order to obtain the
hierarchically small VEV. One should note that the 
condition is not the fine 
tuning of the parameter usually done in the 
quantum field theory. This condition is fulfilled 
by the choice of the flavor set. By minimizing 
the Higgs potential, we obtain that 
\begin{equation}
M_W=M_c~{\rm exp}
\left(-\frac{C^{(4)}}{\abs{C^{(3)}}}~{\rm ln}2 
+\frac{11}{6}\right).
\label{shiki25}
\end{equation}
We see that the large gauge 
hierarchy $M_c\sim M_{GUT}, M_{Planck}$ is realized 
for the large ratio, $C^{(4)}/\abs{C^{(3)}}\gg 1$. 
The magnitude of the ratio for the values of $p=11, 19$, where $p$
is defined by $M_c\equiv 10^p$ 
GeV, is $C^{(4)}/\abs{C^{(3)}}\simeq 32.54~(p=11), 59.12~(p=19)$.
The large gauge hierarchy is realized if 
we have $C^{(2)}=0$ and the large 
ratio $C^{(4)}/\abs{C^{(3)}}$ at the same time.

Let us present a few examples of the flavor set 
in the model I. We choose $(k, m)=(1,0)$ as 
a demonstration. Then, we find that 
$$
(N_{adj}^{(+)},dN_{adj}^{(+)s})=(1, 1), (2, 5),\cdots,~
(N_{fd}^{(+)}, N_{fd}^{(+)s})=(0, 3), (1, 5), \cdots.
$$
\par\noindent
For $(k, m, p)=(1, 0, 19)$, 
$$
(N_{adj}^{(-)}, dN_{adj}^{(-)s})=(0, 29), (1, 33),\cdots,~
(N_{fd}^{(-)}, N_{fd}^{(-)s})=(42, 1), (43, 3), \cdots.
$$
\par\noindent
For $(k, m, p)=(1, 0, 11)$, 
$$
(N_{adj}^{(-)}, dN_{adj}^{(-)s})=(0, 16), (1, 20),\cdots,~
(N_{fd}^{(-)}, N_{fd}^{(-)s})=(22, 1), (23, 3), \cdots.
$$
We observe that the flavor numbers 
$dN_{adj}^{(-)s}, N_{fd}^{(-)}$ are of order $O(10)$. One 
has to take care about the reliability of perturbation 
theory for such the large number of flavor because an expansion 
parameter in the present case may be given by
$(g_4^2/4\pi^2)N_{flavor}$, and it must be 
$(g_4^2/4\pi^2)N_{flavor} \ll 1$ for reliable 
perturbative expansion.

Now, let us study the Higgs mass in the 
model I. The Higgs mass squared is obtained 
by the second derivative of the effective 
potential evaluated at the minimum of the 
potential (\ref{shiki19}),
\begin{equation}
m_H^2
=\frac{3g_4^2}{16\pi^2}~M_W^2~\left(-\frac{C^{(3)}}{6}\right)
\left(=\frac{3g_4^2}{16\pi^2}~M_W^2~k < M_W^2\right).
\label{shiki44}
\end{equation}
The choice $k=1$ is 
the most desirable one for the large gauge 
hierarchy, so that the Higgs mass is lighter 
than $M_W$, which is the same result in the 
original Coleman-Weinberg's paper \cite{cw}. Therefore, one 
concludes that the large gauge hierarchy and the 
sufficiently heavy Higgs mass are not compatible 
in the model I.   
\subsection{Model II}
We study another model called Model II in this 
subsection. We introduce massive bulk 
fermions \cite{takenaga,hty1,marutake} in 
addition to the massless bulk matter in the
model I. We introduce a 
pair of the fields, $\psi_+$ and $\psi_-$ whose 
parity is different to each 
other, $\psi_{\pm}(-y)=\pm \psi_{\pm}(y)$. Then, a parity even 
mass term is constructed like $M{\bar\psi}_{+}\psi_{-}$.

The contribution to the mass term from the 
massive fermions is given by
$$
\frac{1}{2}\zeta(3)C^{(2)}(\pi a)^2
\rightarrow 
-\frac{1}{2}\left[\zeta(3) C^{(2)}
+8N_{pair}B^{(2)}\right](\pi a)^2\quad \mbox{with}
$$
$$
B^{(2)}=\sum_{n=1}^{\infty}\frac{1}{n^3}
\left(1+nz+\frac{n^2z^2}{3}\right)\e^{-nz},
$$
where $N_{pair}$ stands for the number of the 
pair $(\psi^{(+)}, \psi^{(-)})$ and we have 
defined a dimensionless 
parameter $z\equiv 2\pi RM=M/M_c$. We 
observe that the potential is suppressed 
by the Boltzmann-like factor $\e^{-nz}$, reflecting 
the fact that the effective potential shares 
similarity with that in finite temperature 
field theory \cite{dj}.

The essential behavior of the VEV
is governed by the factor $B^{(2)}$, {\it i.e.}
$\pi a_0 \simeq \gamma B^{(2)}$
with some numerical constant $\gamma$ of order $1$.
If we write $\pi a_0 = \e^{-Y}$, then, one 
finds, remembering Eq.(\ref{shiki16}), that
\begin{equation}
-Y={\rm ln}(\pi a_0)\left(={\rm ln}
\left(\frac{M_W}{M_c}\right)\right)=(2-p)
{\rm ln}10 \simeq 
\left\{
\begin{array}{ccc}
-34.539 & \mbox{for} & p=17, \\
-20.723 & \mbox{for} & p=11.
\end{array}\right.
\end{equation}
The gauge hierarchy is controlled by the magnitude 
of $Y$, in other words, the bulk mass 
parameter $z$, and the large gauge hierarchy is 
achieved by $\abs{z}\simeq 30 \sim 40$. The 
large gauge hierarchy is realized by the 
presence of the massive bulk fermion. 
We notice that the flavor number of the massless 
bulk matter is not essential for the large gauge 
hierarchy in the model II.

Now, let us next discuss the Higgs mass 
in the model II. The Higgs mass 
is given by
\begin{equation}
m_H^2=\frac{g_4^2}{16\pi^2}M_W^2
\left[-C^{(3)}{\rm ln}(\pi a_0)+\frac{4}{3}C^{(3)}
+C^{(4)}{\rm ln}2
\right]=\frac{g_4^2}{16\pi^2}M_W^2~F,
\label{shiki56}
\end{equation}
where we have defined 
\begin{equation}
F\equiv -C^{(3)}{\rm ln}(\pi a_0)+\frac{4}{3}C^{(3)}
+C^{(4)}{\rm ln}2.
\label{shiki57}
\end{equation}
The Higgs mass depends on the logarithmic 
factor. We observe that the larger the gauge hierarchy 
is, the heavier the Higgs mass is. An important 
point is that the coefficient $C^{(3)}$ is 
not related with the realization of the 
large gauge hierarchy, so that it is not 
constrained by the requirement of the 
large gauge hierarchy at all.


In order to demonstrate the size of the Higgs mass in the 
model II, let us choose $(k, l, m)=(-4, -1, -1)$. Then, the 
flavor set is given by
\begin{eqnarray*}
(N_{adj}^{(-)}, dN_{adj}^{(-)s})&=&(1, 3),~(2, 7),\cdots,~
(N_{fd}^{(-)}, N_{fd}^{(-)s})=(3, 1),~(4, 3),\cdots,\\
(N_{fd}^{(+)}, N_{fd}^{(+)s})&=&(2, 1),~(3, 3),\cdots,~
(N_{adj}^{(+)}, dN_{adj}^{(+)s})=(1, 0),~(2, 4),\cdots.
\end{eqnarray*}
And the Higgs mass in GeV unit 
is calculated as 
$$
m_H \simeq 
\left\{
\begin{array}{ccc}
119.5 &\mbox{for} & p=17,\\
92.6 & \mbox{for} & p=11,
\end{array}\right.
$$
where we have used $g_4^2\simeq 0.42$. Here, we note that 
in the usual scenario of the gauge-Higgs unification, one 
requires $g_4\sim O(1)$ in order to have the heavy 
enough Higgs mass \cite{ghrelation,hty2}. The large 
gauge hierarchy enhances the Higgs mass sizably 
even for the weak coupling. We observe that for the 
fixed integers $(k, l)$, the large 
gauge hierarchy, that is, large ${\rm ln}(\pi a_0)=-Y$ 
enhances the size of the Higgs mass.
The larger the gauge hierarchy is, the heavier the Higgs 
mass tends to be. 
\section{Conclusions and discussions}
We have studied the five dimensional 
nonsupersymmetric $SU(3)$ model compactified 
on $M^4\times S^1/Z_2$, which is the simplest 
model to realize the scenario of the gauge-Higgs 
unification. We have discussed whether the large 
gauge hierarchy is realized in the scenario or not. 
The Higgs potential is generated at the one-loop 
level and is obtained in a finite form, reflecting 
the nonlocal nature that the Higgs field is the 
Wilson line phase in the gauge-Higgs 
unification. The Higgs potential is calculable 
and accordingly, the Higgs mass, too. 
We have found two models (model I, II), in which the large
gauge hierarchy is realized. The condition $C^{(2)}=0$ is crucial
for our discussions.

In connection with the condition, it may be worth 
mentioning that there are examples, in 
which the loop correction is exhausted at 
the one-loop level (without supersymmetry). They 
are the coefficient of the axial 
anomaly\cite{adler} and the Chern-Simons 
coupling \cite{sakamoto}. As for the latter 
case, a simple reason for the two (higher) loop 
correction not to be generated comes from the 
invariance of the action under the large gauge 
transformation. Since the shift symmetry of the Higgs
potential can be regarded as the invariance under the 
large gauge transformation, one may be able to 
prove that there is no two (higher)-loop correction
to the mass term of the Higgs potential.
In order to confirm it, one needs more studies of 
the higher loop corrections 
to the Higgs potential (mass) in the
gauge-Higgs unification \cite{hoso}.
\begin{center}
{\bf Acknowledgements}
\end{center}  
I would like to thank Professor K. Yamawaki and other members of
the organizing committee for their hospitality and for inviting
me to participate in SCGT 2006.
This work was supported by the 21st Century COE 
Program at Tohoku University.
\vspace*{1cm}

\end{document}